\documentclass[aps,prb,twocolumn,groupedaddress]{revtex4}

\usepackage{float}
\usepackage{amsmath}
\usepackage{amssymb}
\usepackage{siunitx}
\usepackage{chemformula}
\usepackage{graphicx}
\usepackage{dcolumn}
\usepackage{bm}
\usepackage{siunitx}
\usepackage{lmodern}

\usepackage[T1]{fontenc}
\usepackage[utf8]{inputenc}
\usepackage{textcomp}
\usepackage{epstopdf}
\usepackage{natbib} 
\usepackage{hyperref}
\usepackage{ulem}
\makeatletter
\begin{document}

\title{Evolution of amorphous structure under irradiation: zircon case study}
\author{A. Diver, O. Dicks}
\affiliation{School of Physics and Astronomy, Queen Mary University of London School of Physics and Astronomy, Mile End Road, London, E1 4NS, UK}
\author{A. M. Elena, I. T. Todorov}
\affiliation{Daresbury Laboratory STFC UKRI, Scientific Computing Department, Keckwick Lane, Daresbury WA4 4AD, Cheshire, UK}
\author{K. Trachenko}
\affiliation{School of Physics and Astronomy, Queen Mary University of London School of Physics and Astronomy, Mile End Road, London, E1 4NS, UK}

\begin{abstract}
The nature of the amorphous state has been notably difficult to ascertain at the microscopic level. In addition to the fundamental importance of understanding the amorphous state, potential changes to amorphous structures as a result of radiation damage have direct implications for the pressing problem of nuclear waste encapsulation. Here, we develop new methods to identify and quantify the damage produced by high-energy collision cascades that are applicable to amorphous structures and perform large-scale molecular dynamics simulations of high-energy collision cascades in a model zircon system. We find that, whereas the averaged probes of order such as pair distribution function do not indicate structural changes, local coordination analysis shows that the amorphous structure substantially evolves due to radiation damage. Our analysis shows a correlation between the local structural changes and enthalpy. Important implications for the long-term storage of nuclear waste follow from our detection of significant local density inhomogeneities. Although we do not reach the point of convergence where the changes of the amorphous structure saturate, our results imply that the nature of this new converged amorphous state will be of substantial interest in future experimental and modelling work.
\end{abstract}

\maketitle

\section{Introduction}

Research into materials suitable for the encapsulation of nuclear waste is both a societal and scientific challenge. Part of this challenge is how to deal with existing stockpiles of plutonium. Currently the UK has the largest civilian plutonium inventory in the world, expected to be 140 tonnes this year \cite{HYATT2017303}. Two solutions being considered are reprocessing Pu into mixed oxide fuel or immobilization by encapsulation in a waste form followed by deep geological disposal. 
 
Many materials have been proposed as potential immobilization matrices (waste forms), but no matter how resistant any material is to radiation damage after a few thousand years \cite{weber_ewing_catlow} any initial crystalline waste form with a typical waste loading will undergo a transformation, becoming completely amorphized long before it is deemed safe. In fact, it is typically considered that waste forms need to be an effective barrier for long-lived nuclear waste encapsulation for up to a 1,000,000 years. For a model waste form, zircon, this transition to a fully amorphous state has been measured experimentally \cite{weber_ewing_catlow} and occurs at doses between \num{e18} and \num{e19} alpha decay events/g, which corresponds to a waste storage time in the order of 1,700 years.
 
Previous modeling work has only simulated high-energy radiation damage effects in the crystalline phases of materials, which only accounts for a small fraction of the waste form’s life cycle. Herein lies the key issue; it cannot be assumed that the amorphous state of zircon is itself unchanging when exposed to radiation (a temptation for such as an assumption lies in considering that introducing more damage does not increase disorder in an already amorphous system). Therefore, it is necessary to study radiation damage effects in the long-term amorphous phase in order to understand the long-term performance of the waste form.

In addition to nuclear waste encapsulation, the question of the evolution of amorphous structure is of fundamental importance. Understanding the amorphous state and ascertaining its main characteristics in the short and medium range order has a long history (see, e.g., Refs. \cite{zallen,wright2013} for review) and is developing as new high-quality data come in. Understanding the ``problem of glass'' was interestingly regarded as the most profound problem in modern physics \cite{anderson}. It is remarkable that despite more than one hundred years of modern research, there remain fundamental open questions regarding the nature of the amorphous state. For example, although the ``continuous random network'' structure proposed in early days of glass science is consistent with the wealth of experimental data, the presence of quasi-crystalline cybotactic groups can not be ruled out by the current experimental data \cite{wright2013}. 

Here, we pose a fundamental question of whether an amorphous structure produced by liquid quench (glass) can be made ``more amorphous'' in a sense of reducing the degree of order in the short- and medium range? We also explore related and more specific questions: how does the amorphous structure evolve during progressive introduction of radiation damage, does the damaged structure converge to a distinct amorphous state with new short- and medium range order and what is the nature of this new amorphous state?

The work done on distinguishing in which range crystalline and amorphous states differ \cite{wrightobservations} can be built upon to aid this endeavour. Wright suggests that to answer these question we need to look at the composition of structural units, their connectivity and fluctuations in the local density of the glass to determine the differences in overall network topology between crystalline and amorphous states. The same observables can also be used to distinguish between evolving amorphous states in a similar fashion.

In this paper, we address these questions using molecular dynamics (MD) simulations. We use zircon (ZrSiO$_4$) as a case study, for several reasons. First, it ``accepts'' all damage and shows very little recovery to the crystalline state. Second, it is a well characterized. Third, it was considered as a potential waste form for the encapsulation of plutonium because of its high durability \cite{ewing_lutze_weber_1995}. 

Radiation damage experiments in zircon have also been conducted, including analyzing naturally radiation damaged zircon samples \cite{murakami1991alpha}, Pu doped samples over periods of 6.5 years \cite{weber_ewing_wang_1994} and heavy ion beam experiments involving a range of different ions and energies \cite{weber_ewing_wang_1994}. Some work has been carried out simulating amorphous zircon structures \cite{osti_963199}. These simulations show Zr and Si become under-coordinated, with a corresponding volume expansion, when compared to the crystalline phase. Previous radiation damage simulations in crystalline zircon have focused on structural effects like polymerization of SiO$_n$ units and percolation \cite{PhysRevB.65.180102,Trachenko_2004per}. However, there have been no high energy collision cascades been performed in amorphous zircon.

Here, we perform large-scale MD simulations of high-energy overlapping collision cascades in both crystalline and amorphous zircon in order to ascertain and understand the difference in response of the two structures and develop a new algorithm to detect and quantify the radiation-induced defects. We find that an average property such as pair distribution function does not detect the differences between the damaged configurations in the amorphous structures. On the other hand, our new analysis method based on local coordinations finds that the amorphous structure substantially {\it evolves} due to radiation damage. Our analysis shows a notable correlation between the local structural changes and enthalpy. We also find significant local density inhomogeneities in the amorphous structure due to radiation damage, which have important implications for the long-term encapsulation of nuclear waste. Although we do not reach the point of convergence where the changes of the amorphous structure saturate, our results open up the way to study this process and imply that ascertaining the nature of this new converged amorphous state will be of substantial interest in future experimental and modelling work.

\section{Methods}

\subsection{Molecular dynamics simulation method}

The molecular dynamics package  \texttt{DL\char`_POLY\char`_4} , which uses domain decomposition to simulate systems with large numbers of atoms in parallel \cite{todorov_smith_trachenko_dove_2006}, has been used to simulate the propagation of heavy recoils due to alpha decay of Pu in crystalline and amorphous zircon. We model radiation damage to the heavy recoil nucleus of an alpha decay, specifically U atoms with a 70 keV recoil energy. During the collision cascade thousands of atoms are displaced in the system, with the permanently displaced atoms causing amorphization of the material. In order to analyze the extent of these atomic displacements and their effect on the material structure this work makes use of new on-the-fly analytical tools we have developed for \texttt{DL\char`_POLY\char`_4.10} (pre-release), including time resolved coordination analysis. All collision cascades were carried out using Archer high-performance computing facility. Each cascade was simulated in a box with 10125000 atoms using 1152 processors. Each simulation took about 20 hours and resulted in 180 ps of trajectory time. 

Interatomic potentials that have been previously used and well tested for zircon \cite{Trachenko_2004per} have been used in this study. The Zr-O and O-O interactions are modeled using Buckingham potentials, and for the Si-O interactions a Morse potential is used. These pairwise potentials are then joined at a short-range with a repulsive ZBL (Ziegler Biersack Littmark) \cite{ziegler_biersack_1985} stopping potential. Simulation boxes containing in the order of 10 million atoms were used for both the amorphous and crystalline structures. These large cell sizes were necessary to contain multiple overlapping collision cascades. All collision were carried out using the NVE ensemble. A variable timestep was used in these simulations, starting at 1 fs, in order to account for the initial high velocities of particles.

In order to study the effect of multiple cascades overlapping, each U recoil ion was placed at the same initial starting coordinates within the simulation box and given an impulse along the same initial trajectory, with the only difference being the initial structure being taken from the previous cascade. This was done in order to maximise the damage in one area of the simulation box and to observe the effects of damage saturation on the local structure.

\subsection{Producing amorphous zircon}

The amorphous structures of zircon were produced by a slow melt quenching of the same crystal structures used in this work. First the crystal structure was melted at 6500 K for 100 ps using an NPT ensemble and then subsequently quenched to 300 K using a quench rate of 10  \si{K.ps^{-1}} . This quench rate is much higher than those used in experiments, but a slower quench rate of 1 \si{K.ps^{-1}} on a smaller system of 100,000 atoms resulted in little difference in overall structure. The volume of the amorphous structure produced by the melt quench was calculated to be 16\% greater than the crystalline zircon which compares well with the experimentally measured volume increase due to amorphization in zircon of 18\% \cite{weber_ewing_wang_1994}.

We note that as the motivation behind creating the amorphous structures comes from the amorphization of zircon due to collision cascades one might ask why the amorphous structures were not produced by collision cascades. Each individual alpha decay event produces in the order of a few thousand permanently displaced atoms, as shown from nuclear magnetic resonance (NMR) studies of natural radiation damaged zircons \cite{farnan_salje_2001} and our own work.  This means that the amount of collision cascades necessary to completely amorphize zircon via molecular dynamics would be around 2,000 MD simulations, if we assume approximately 5,000 atoms are displaced per collision cascade. Therefore, the amount of time, and associated computing cost, for achieving an amorphous state via simulated cascades is not practical.

\section{Results and analysis}

\subsection{Defining damage in an amorphous structure}

Previous works quantify the effects of radiation damage cascades by measuring the number of displaced or ‘defect’ atoms \cite{PhysRevB.73.174207}. An atom is considered displaced if it moves more than a set cut off radius away from its initial position. Some of these displaced atoms may move to crystalline positions, resulting in damage recovery. Therefore, a defect is considered to occur if there is either no atom within a sphere of a given radius centred on an initial crystalline atomic position (a vacancy defect) or if there is more than one atom inside that same sphere (an interstitial defect). Both definitions of displacement and defect rely on the initial atomic positions, which is reasonable for crystalline systems but is inadequate for defining defects in glass or amorphous systems. Indeed, in a crystal the atoms in the surrounding lattice to a displaced atom relax back to their crystalline positions, with small perturbations, and therefore are not counted as defects. However, in an amorphous system, if a bond is broken in a glass forming unit, such as a tetrahedron or octahedron, and an atom is displaced, then the surrounding glass forming units may relax back to positions displaced from their initial coordinates, whilst maintaining the same connectivity of the glass network. To ensure that only the truly ``displaced'' atoms are counted, and not all of the  atoms in the surrounding polyhedra, a new definition for defects and displaced atoms must be considered, using the connectivity of the atoms rather than their initial positions, due to the lack of crystalline coordinates.

Here, we develop a different method which measures the change in local coordination of each atom due to the collision cascades. The nearest neighbour coordination of an atom is defined by cutoffs between different pairs of atomic species. In this paper the cutoff radius is defined as the first minimum after the first peak in the individual pair distribution functions, g$_{ij}$(r), of each pair of species.

Our on-the-fly analysis allows two different coordination based quantities to be measured. The first one can be used to track displaced atoms during a cascade. We define an atom to be displaced if there is (a) a change to the specific atoms it is coordinated to, and (b) it is displaced by more than a cutoff distance from its initial position. A Si atom that remains bonded to the same O atoms that form its tetrahedral cage is not considered a displaced atom, regardless of any translation, as it does not lead to a change in the glass network. However, if the O atoms the Si is coordinated with are replaced by different O atoms, even if the coordination number is maintained, it is considered displaced if it has also moved a minimum distance from its initial position, as the glass connectivity has changed.

The second quantifiable measure is the overall distribution of species-species coordination numbers, and how they change as a result of radiation damage. The coordination distribution of each species A-B pair is calculated every $n$ time steps, and can be plotted as a function of time. This can be used to measure the change in the overall number of 'defects' in the system before, during, and after a cascade. The average change in 'defect' number as the result of a radiation event is also calculated in this paper. The initial coordination distribution average is calculated during a simulation run of the structure before any collision cascades and then compared to the average of each coordination distribution over the last 20 ps of the cascade simulation run.

Both of these methods can be used to monitor time resolved changes to the structure of the material during a cascade, and the overall changes to the structure as a result of a single, or multiple cascades.

\subsection{Collision cascades}
\subsubsection{Pair distribution functions in both crystalline and amorphous zircon after collision cascades}

\begin{figure}
\centering
\includegraphics[width = 8cm]{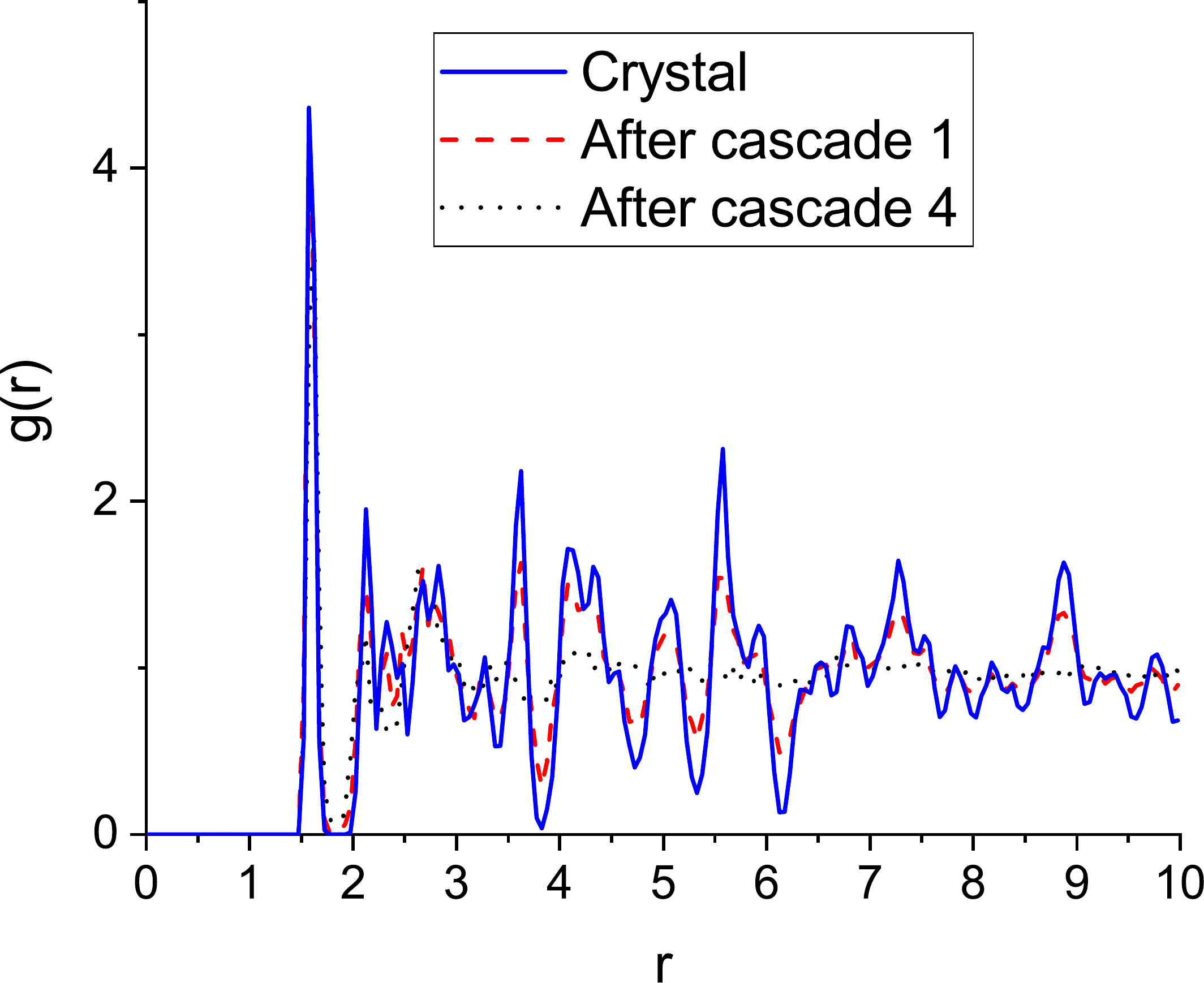}
\caption{The total radial distribution function of crystalline zircon in the 'overlap' region near the initial recoil site, before any cascades, and after the 1\textsuperscript{st} and 4\textsuperscript{th} cascade.}
\label{fig:totalg(r)crystal}
\end{figure}

\begin{figure}
\centering
\includegraphics[width = 8cm]{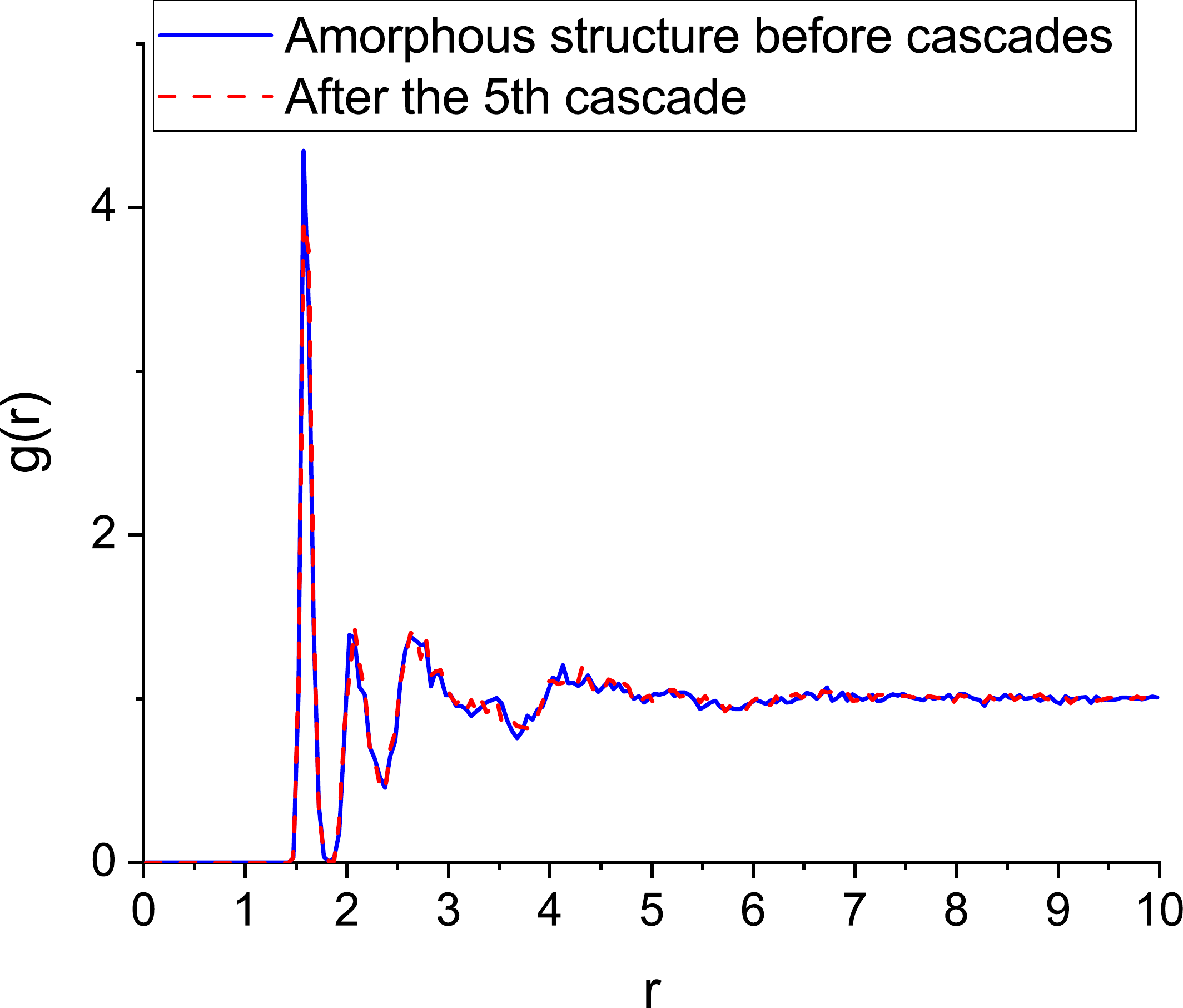}
\caption{Total radial distribution function for amorphous zircon in the overlapping region before and after cascade 5 in the region of the cascades}
\label{fig:totalg(r)amorphous}
\end{figure}

We firstly analyze the pair distribution functions (PDF) of the crystalline and amorphous structures damaged by collision cascades.
The PDFs of the total systems (averaged over ~10 million atoms) are insensitive to radiation damage, there is no discernible change to the crystalline or amorphous PDF before and after the collision cascades as the structural changes are small when compared to the total system size. 

Some use can be made of the PDF if the regions of maximum damage are zoomed in on. The PDF was calculated in the structure containing around 6000 atoms and encompassing the collision cascades. In Fig. \ref{fig:totalg(r)crystal} we calculate the PDF peaks of a region of approximately 6000 atoms affected by the collision cascade. It can be observed that the peaks in the PDF of crystalline zircon decrease and subsequently disappear as more collision cascades overlap. This is characteristic of the crystalline-amorphous transition where only short and medium-range order is present in the amorphous state. However, the effect of radiation damage in the amorphous structure in a similarly sized region (see Fig. \ref{fig:totalg(r)amorphous}) is markedly smaller: the only observable difference is a decrease in the height of the first peak of the PDF.

The insensitivity of the PDF to local structural changes reinforces earlier discussion \cite{wright2013,wrightobservations}, and underscores the need for more sensitive and local analysis. This will become more apparent in our analysis of coordination changes and local density changes discussed later.

\subsubsection{Coordination statistics in crystalline zircon due to multiple cascades}

\begin{figure}
\centering
\includegraphics[width = 8cm]{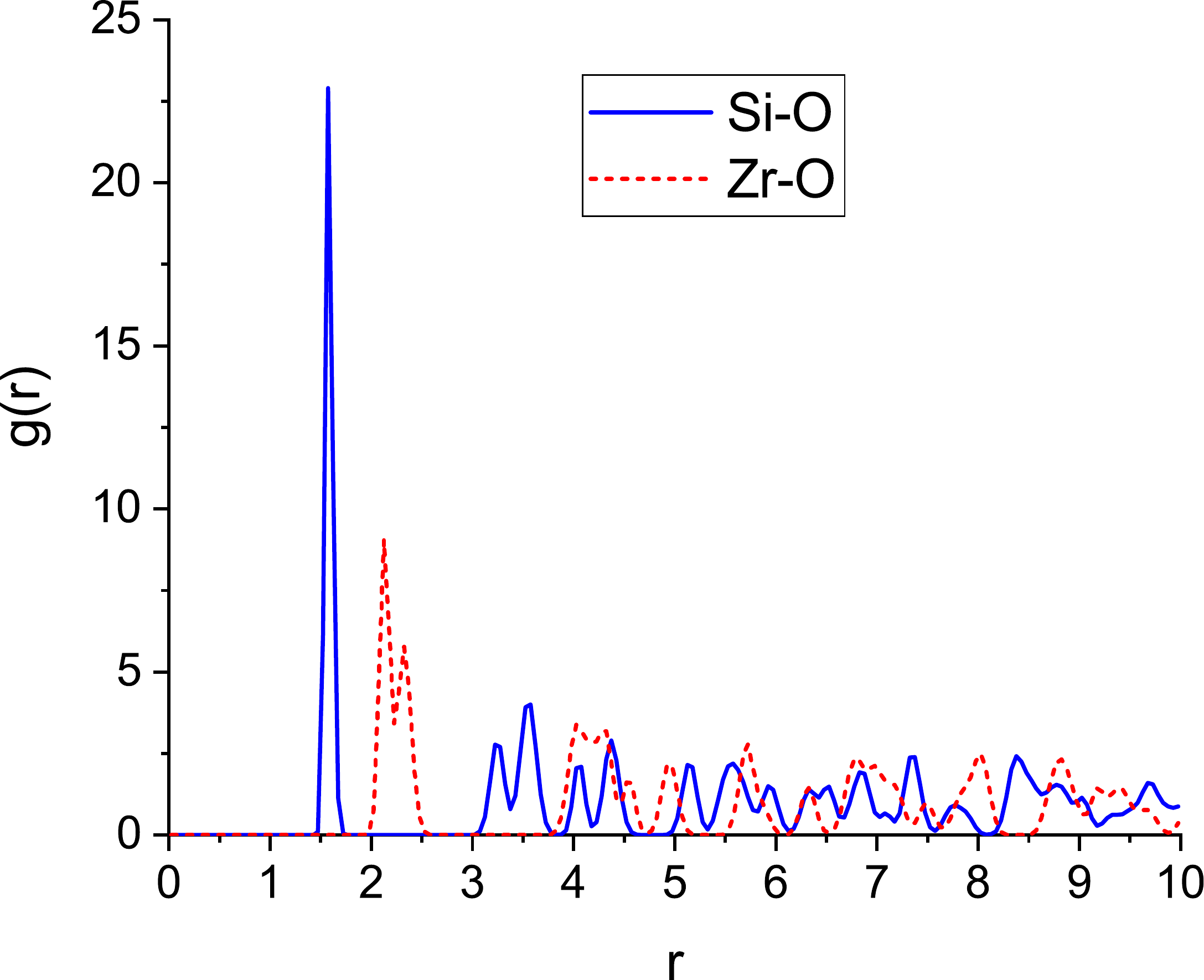}
\caption{Pair distribution functions, $g_{ij}(r)$, of Si-O and Zr-O pairs in crystalline zircon before any radiation damage.}
\label{fig:Zr-Ogr}
\end{figure}

For all collision cascades in the crystal structure the cut offs to build local atomic coordination are 2.0 \si{\angstrom} and 2.80 \si{\angstrom} for Si-O and Zr-O respectively, with these values taken from the first minimum after the first maximum of the pair distribution functions in Fig. \ref{fig:Zr-Ogr}. These values are then used in our coordination displacement definition with the specific atom needing to also be displaced by a minimum 0.75 \si{\angstrom}, 1.0 \si{\angstrom} and 1.0 \si{\angstrom} respectively for Si, O and Zr atoms.  The values for the minimum displacement come from being slightly less than half the bond length between atoms. For O we choose the largest bond length pair to define this.

\begin{figure}
\centering
\includegraphics[width = 6cm]{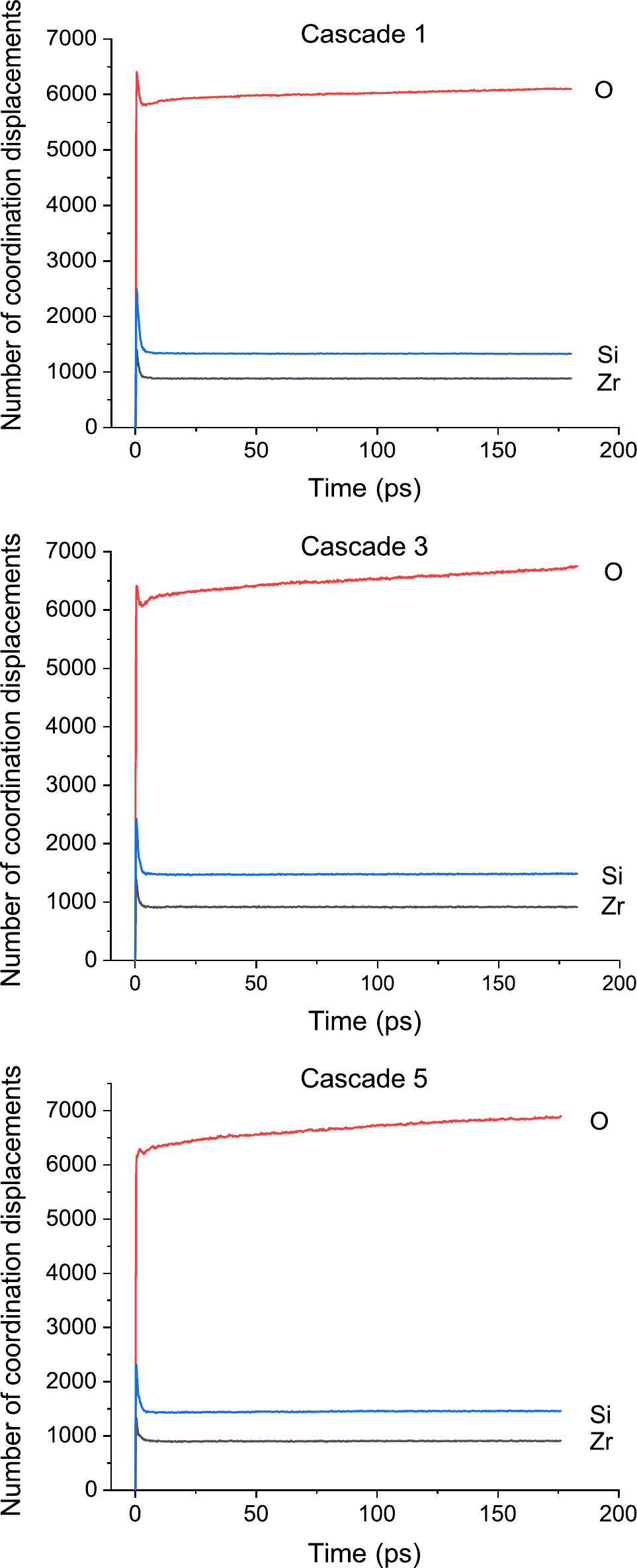}
\caption{The time evolution of coordination displacements of Zr, O and Si during collision cascades 1, 3 and 5 in crystalline zircon.}
\label{fig:coordchangecrystal}
\end{figure}

In Fig. \ref{fig:coordchangecrystal} the evolution of coordination displacements during the initial cascade simulation (cascade 1) and subsequent cascades is shown. We observe that overlapping cascades produce a similar number of displacements to the previous cascades, and that the displacements per recoil do not increase when a damaged region undergoes further radiation damage. This result is at odds with previous work on radiation damage in crystalline zirconolite \cite{Chappell_2012}, where a second overlapping cascade leads to an increase in displaced atoms in the already damaged region. However, there are many factors which could explain this discrepancy, the two different materials may have different radiation resistance properties especially as zirconolite contains more mobile alkali ions. The previous study also used simulations cells approximately a 10\textsuperscript{th} of the size of the simulation cells used here, which may have lead to larger temperature increases during the cascade. 

Although each overlapping cascade produces similar numbers of total displacements, there is a change in how the system relaxes from the peak number of displacements in the first 5 ps after the recoil (see Fig. \ref{fig:coordchangecrystal}). Previous work \cite{farnan2007} looking at naturally damaged zircon samples estimated around 830 Si atoms are displaced per alpha decay event which is in good agreement with the number of displaced Si atoms we get after the first cascade being around 1300.
As more cascades overlap the amount of recovery of displaced O atoms decreases and becomes insignificant by the 5\textsuperscript{th} cascade, meaning once a region becomes significantly damaged less recovery occurs due to amorphization, and atoms are unable to return to their crystalline lattice positions. 

\begin{figure}
\centering
\includegraphics[width = 8cm]{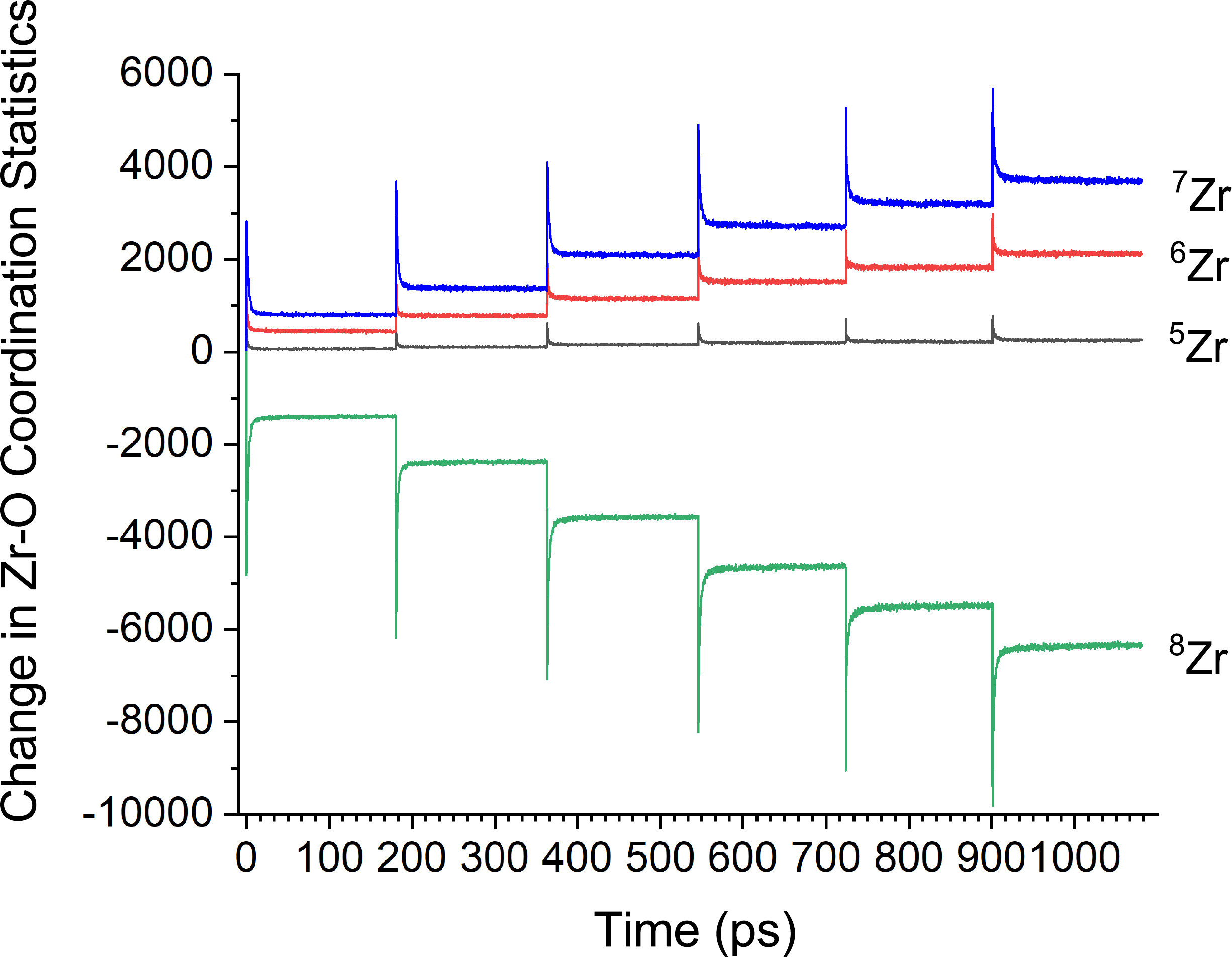}
\caption{The change in Zr-O coordination (where \textsuperscript{5}Zr is 5-coordinated Zr-O) during consecutive overlapping U recoil cascades in crystalline zircon, as a function of time.}
\label{fig:Zr-Ocoordstats}
\end{figure}

\begin{figure}
\centering
\includegraphics[width = 8cm]{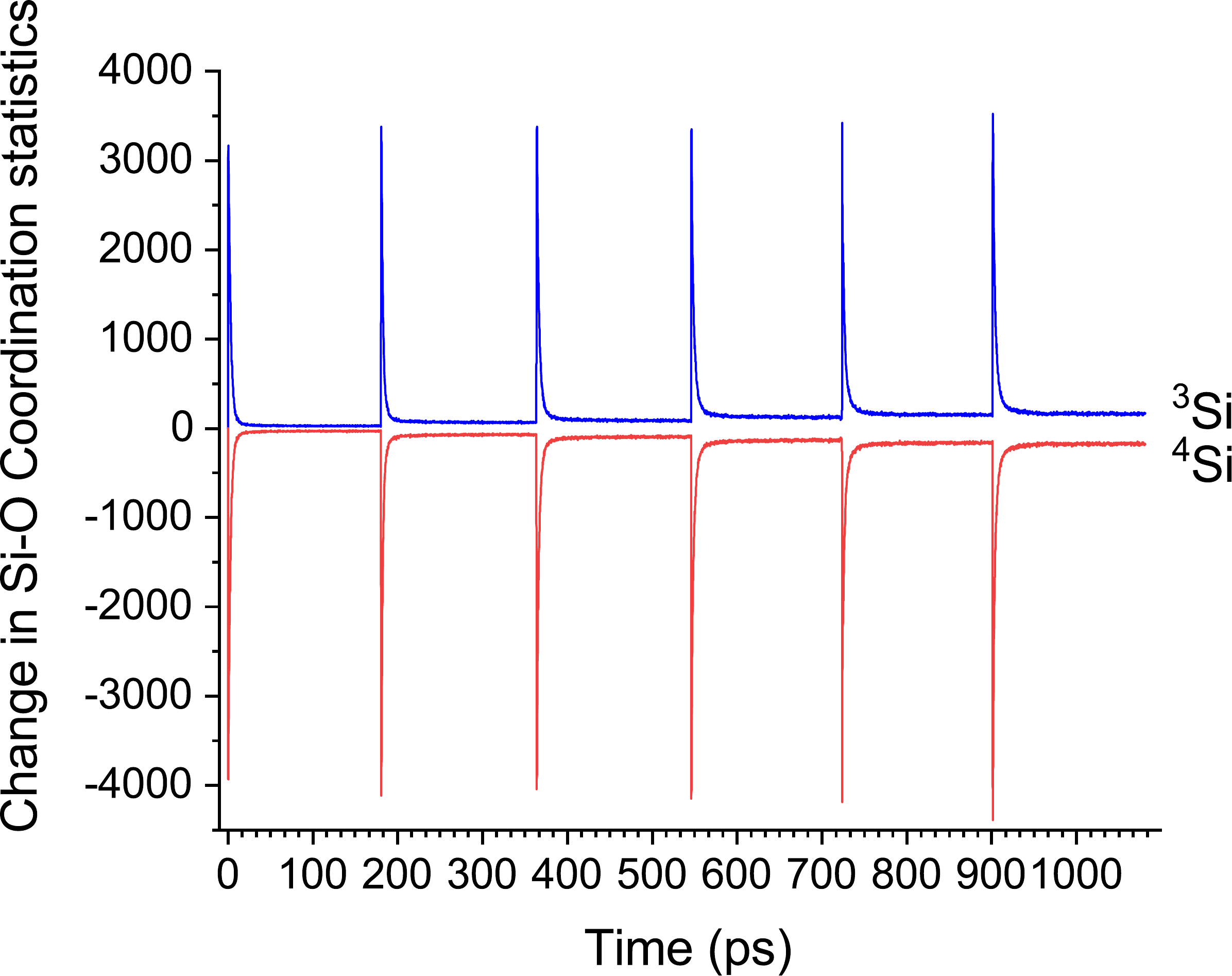}
\caption{Change in Si-O coordination during consecutive collision cascades in crystalline zircon.}
\label{fig:Si-Ocoordstats}
\end{figure}

To see a measurable difference between each cascade, and to quantify changes to the overall microscopic structure, we follow the change in overall coordination statistics as seen in Fig. \ref{fig:Zr-Ocoordstats} and Fig. \ref{fig:Si-Ocoordstats}. Both the Zr-O and Si-O coordination changes reach a steady equilibrium within 20 ps of the recoil after an initial spike. It is interesting to note that the number of displaced atoms recovers to a new equilibrium in much less time, within 1 ps, but the recovery is much smaller proportionally. This suggests that while approximately 10,000 atoms become displaced during a cascade, in the system fewer than 1,500 atoms change their local coordination, or become topological `defects'.

It is also possible to compare the difference between the number of Si atoms that are displaced from Fig. \ref{fig:coordchangecrystal} and the overall change in Si-O coordination from Fig. \ref{fig:Si-Ocoordstats}. After the initial cascade there are roughly 1,200 Si atoms that can be considered to be `displaced', but there is only an increase of 30 3-coordinated Si atoms from the original 4.  Even after 3 cascades there is only a total change of 120 3-coordinated Si atoms due to the damage, which is orders of magnitude lower than the damage suggested by only examining the number of displacements, which is regularly reported. Surprisingly the opposite effect is seen in Fig. \ref{fig:Zr-Ocoordstats} where the amount of coordination changes is greater than the number of displaced Zr atoms as shown in Fig. \ref{fig:coordchangecrystal}. The number of Zr displacements is fewer than the number of coordination changes. This is likely to be because Zr will undergo a change in coordination when O ions are displaced, whilst the much heavier and larger Zr ion itself does not move from its original equilibrium position in the cation sublattice and is therefore not treated as a displacement by our algorithm. O can be considered to be more mobile than Si and Zr. It can also be observed that radiation damage decreases the number of 8-coordinated Zr in the crystal with each subsequent overlapping cascade, as the damaged region becomes amorphized and the local average coordination decreases.

\subsubsection{Coordination statistics in amorphous zircon due to multiple overlapping cascades}

\begin{figure}
\centering
\includegraphics[width = 6cm]{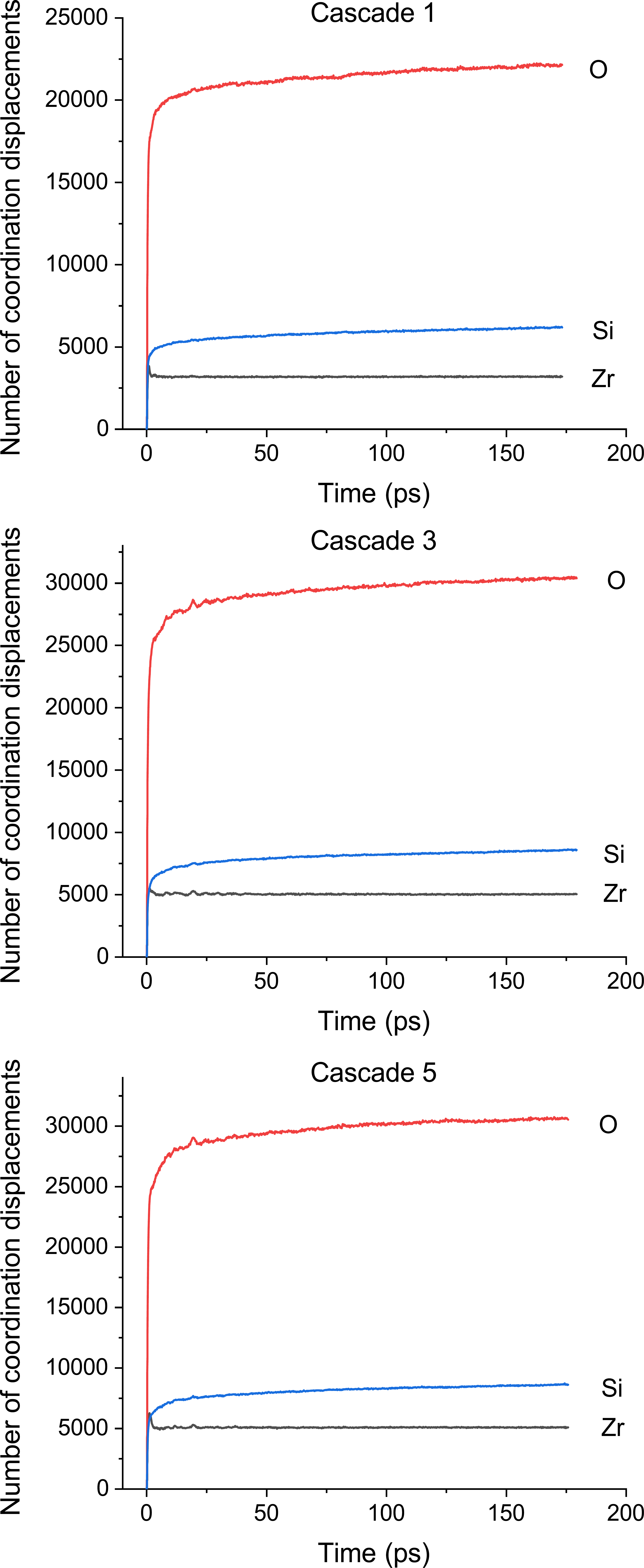}
\caption{Evolution of coordination displacements of Zr, O, Si during cascades 1,3,5 in an initial amorphous zircon structure.}
\label{fig:coordchangamorphous}
\end{figure}

\begin{figure}
\centering
\includegraphics[width = 8cm]{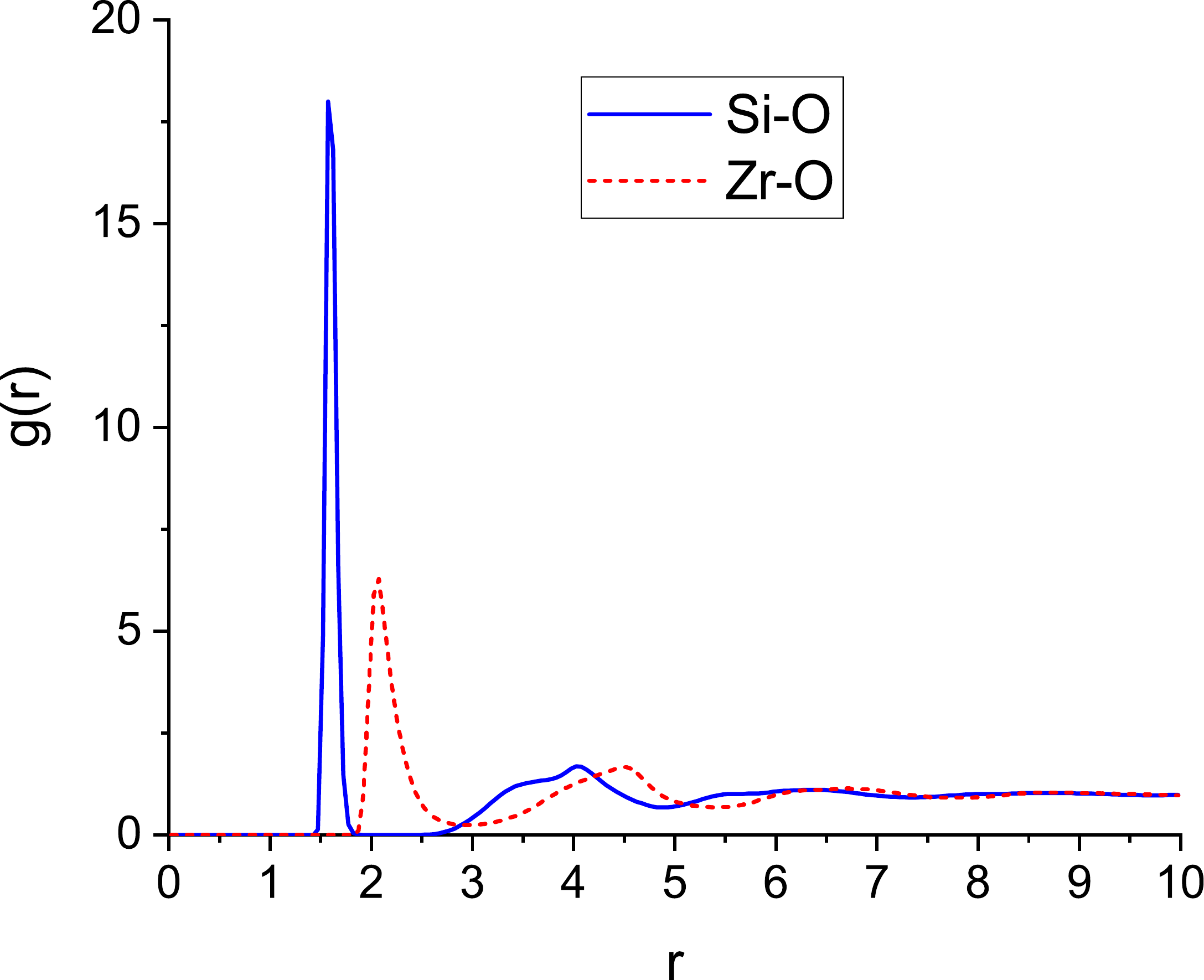}
\caption{Pair distribution functions, $g_{ij}(r)$, of Si-O and Zr-O pairs in amorphous zircon before any radiation damage.}
\label{fig:Si-Ogramorphous}
\end{figure}

In this section, we analyze the radiation damage in the amorphous zircon structure. The change in local atomic coordination in the amorphous zircon system is shown in Fig. \ref{fig:coordchangamorphous}, with the cut offs chosen to be 2.0 \si{\angstrom} and 2.96  \si{\angstrom} for Si-O and Zr-O respectively, where once again these values are taken from Fig. \ref{fig:Si-Ogramorphous}. To be defined as a displacement, the minimum distances the atom must move, as well as undergoing a coordination change, are 0.75 \si{\angstrom}, 1.0 \si{\angstrom} and 1.0 \si{\angstrom} respectively for Si, O and Zr atoms. These were the same values as used in the crystalline case for direct comparison.

Interestingly, there is no sharp recovery peak at 1 ps in the amorphous structure Fig.\ref{fig:coordchangamorphous}, in contrast to crystalline zircon. In fact it is very similar to how the 5\textsuperscript{th} cascade in the crystal shows no peak in the O displacements, because the region of the cascade propagation is already amorphized. In crystalline materials, the sharp recovery peak was attributed to reversible elastic deformation of the structure surrounding the collision cascade: the strongly-anharmonic motion in the cascade core results in the volume expansion (similar to thermal expansion due to anharmonicity) which exerts tensile stress on the surrounding lattice \cite{PhysRevB.73.174207}. The associated strain displaces atoms in the surrounding lattice, resulting in the brief increase of displaced and defect atoms around 1 ps. This displacement is elastic and reversible, resulting in a sharp peak. The observed absence of the peak in amorphous structure interestingly implies that the response of the amorphous structure to radiation damage is more slack in the above sense of reversible elastic deformation.

Another notable observation is that the peak number of coordination displacements produced by cascades in the amorphous system is approximately 4 times greater than that in the crystalline phase and shows an increase with cascade number from the initial cascade to the 3\textsuperscript{rd} cascade. Hence the amorphous structure is softer, atoms are more easily displaced and it is more susceptible to damage. However from the 3\textsuperscript{rd} cascade onwards this increase in the peak displacements disappears, which may be due to the amorphous system produced by melt-quenching differing from one produced by radiation damage. Therefore, saturation in the number of displaced atoms in the heavily damaged structure by multiple overlapping collision cascades is reached. 

\begin{figure}
\centering
\includegraphics[width = 8cm]{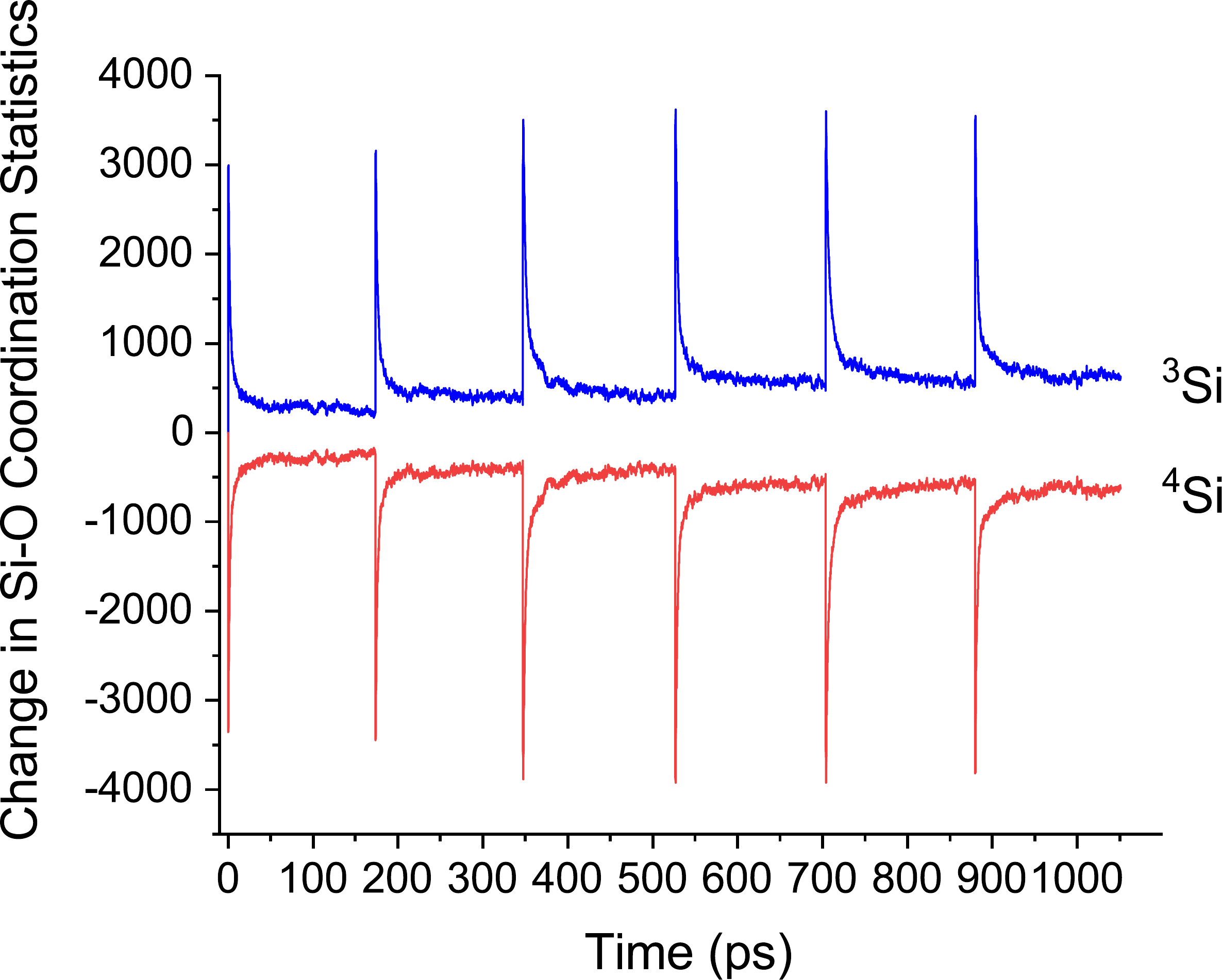}
\caption{Change in Si-O coordination during overlapping collision cascades in amorphous zircon.}
\label{fig:Si-Ocoordstatsamorphous}
\end{figure}

\begin{figure}
\centering
\includegraphics[width = 8cm]{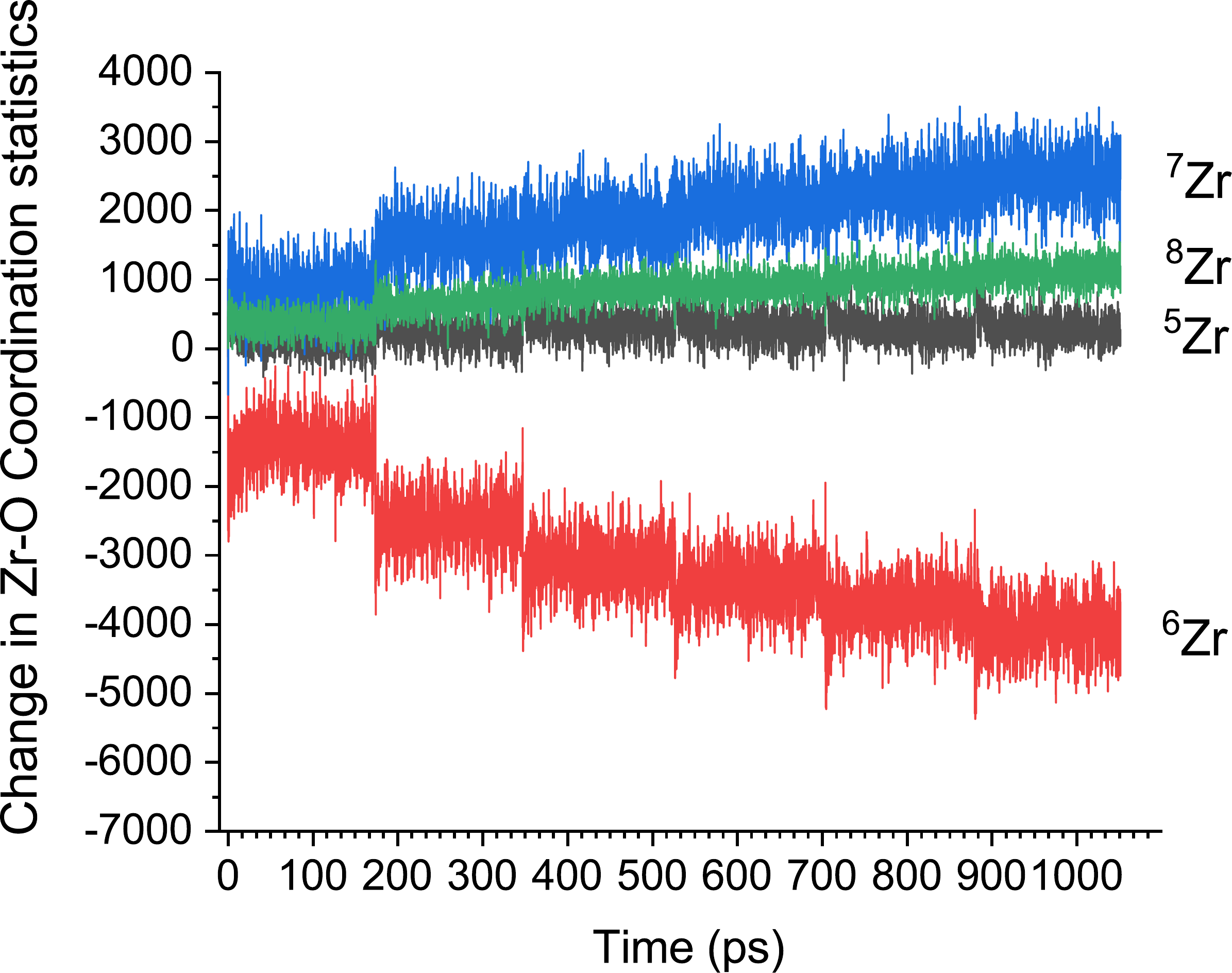}
\caption{Change in Zr-O coordination during collision cascades in amorphous zircon.}
\label{fig:Zr-Ocoordstatsamorphous}
\end{figure}

Although the number of displaced atoms per collision saturates, local structural changes still evolve. Indeed, measurable differences of local structural changes can be analyzed by observing the overall change in coordination statistics as shown in Fig. \ref{fig:Si-Ocoordstatsamorphous} and Fig. \ref{fig:Zr-Ocoordstatsamorphous}. Again, as with the crystalline case, the number of Si atoms identified as displaced atoms is much higher, roughly 6,000, but the number of newly created 3-coordinated Si-O is only around 200. The majority of displaced Si atoms relax back into a tetrahedral configuration, merely changing their nearest neighbour O atoms. After 5 cascades only 600 3-coordinated Si-O are produced, orders of magnitude fewer than the number displaced.

There is a greater degree of fluctuation in the Zr-O coordination number, with noise between steps predominantly due to the minimum of the g(r) not going to zero after the first peak. Although these fluctuations might seem rather large, they account for less than 0.03\% of all Zr atoms inside the MD box. The fluctuation of coordination numbers is due to vibrational motion causing some atoms to move in and out of the cutoff radius, thus switching their coordination. However, it can still be clearly seen that, despite the large variations in coordination number during each run, the cascades lead to a substantial change in the average coordination number. The long simulation times (180 ps) show that the average is stable after each individual cascade.

An interesting trend in the data is the decrease of 6 coordinated Zr-O accompanied by an increase in 7 and 8 coordinated Zr-O after each cascade. The decrease in 6 coordinated Zr suggests there may be a saturation limit for the number of coordination changes produced by a collision cascade. As crystalline zircon undergoes a decrease in average Zr-O coordination, and the melt-quench amorphous zircon undergoes an increase, we hypothesize there is a stable coordination distribution between them, where the structure is healed as quickly as it is damaged by radiation. This hypothesis is further supported by the change in 6 coordinated Zr-O in the amorphous system after each cascade, decreasing from 1500 initially to roughly 500 from the third cascade onwards. This suggests a possible saturation limit is being approached where fewer coordination changes can occur and remain stable after collision cascades.

\begin{figure}
\centering
\includegraphics[width = 8cm]{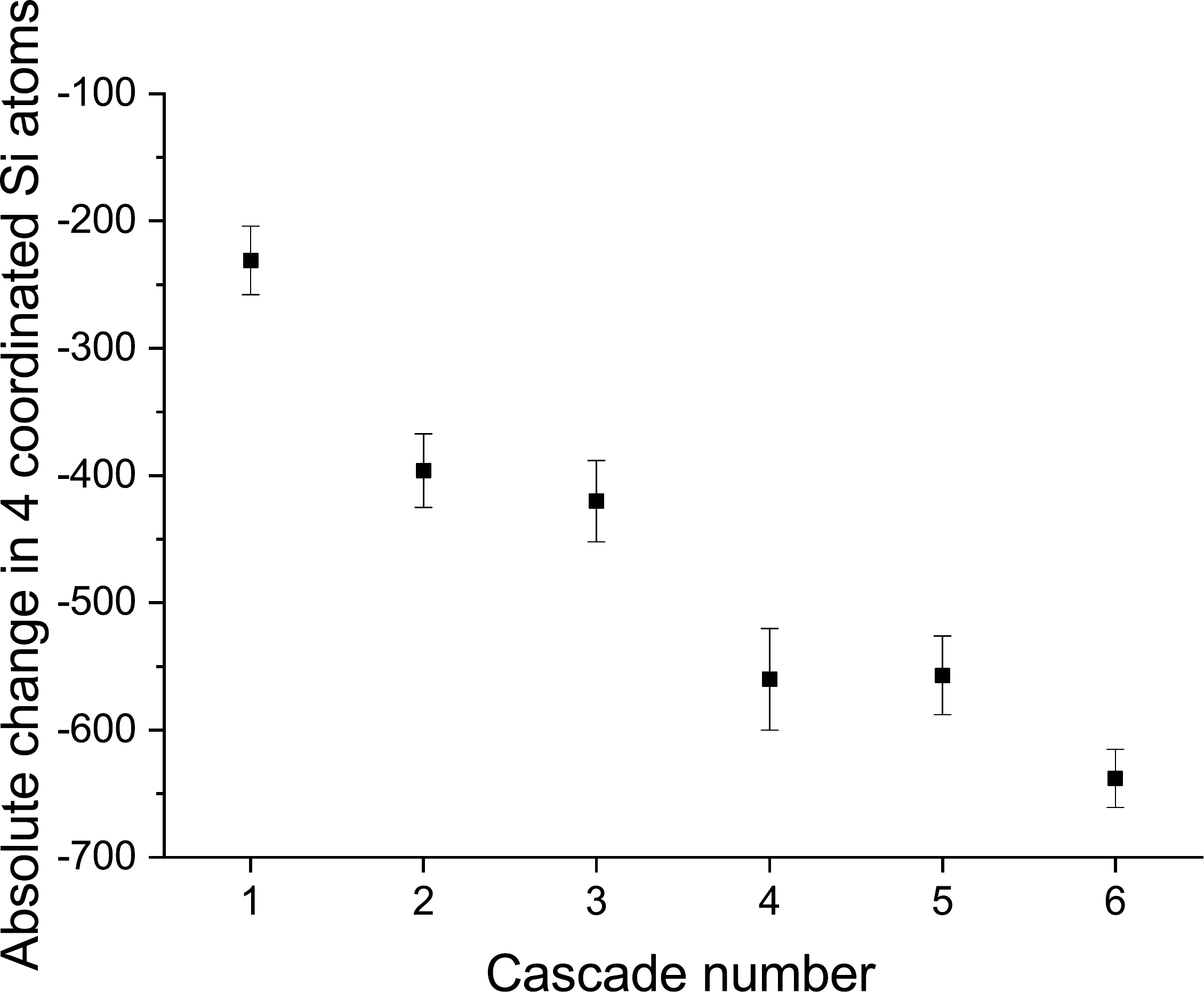}
\caption{Total decrease in number of 4 coordinated Si-O compared to the initial structure after each collision cascade in amorphous zircon.}
\label{fig:abs4Si-O}
\end{figure}

The same effect is seen with a decrease in the number of 4-coordinated Si-O after each cascade, both in the absolute change in number of 4-coordinated Si from the original structure (Fig. \ref{fig:abs4Si-O}) and the change between subsequent cascades. 

\subsubsection{Other structural and energetic effects}

As well as examining the effects of the recoil nucleus on atomic displacements, changes to other structural and macroscopic properties, like enthalpy, can be studied. 

\begin{figure}
\centering
\includegraphics[width = 8cm]{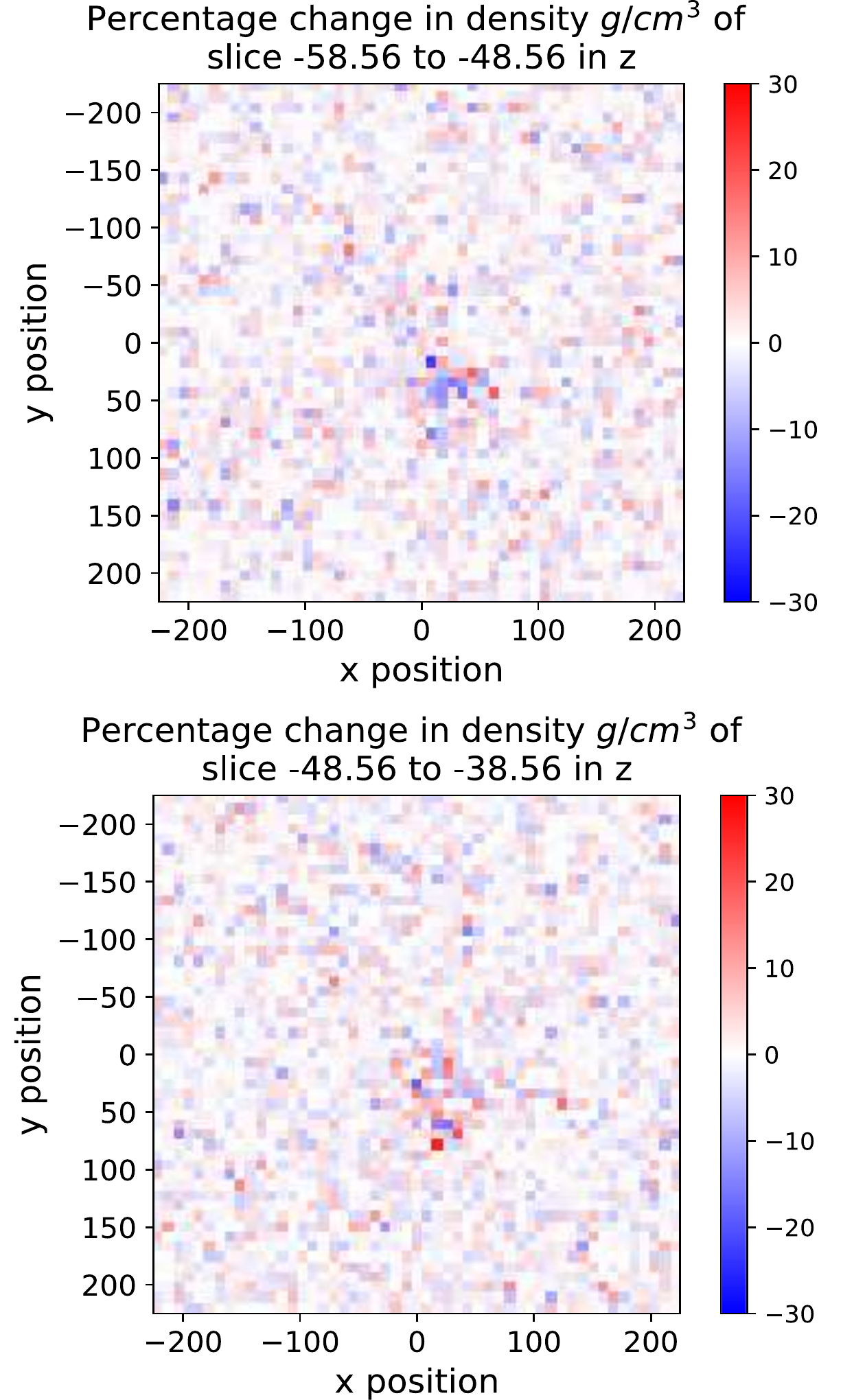}
\caption{The change in density of amorphous zircon after the 1\textsuperscript{st} collision cascade. As shown by the scale, the more red a region, the larger the increase in percentage change in density and the bluer a region the larger the decrease in percentage change in density.} 
\label{fig:amorphousdensityafteroverlap1}
\end{figure}

\begin{figure}
\centering
\includegraphics[width = 8cm]{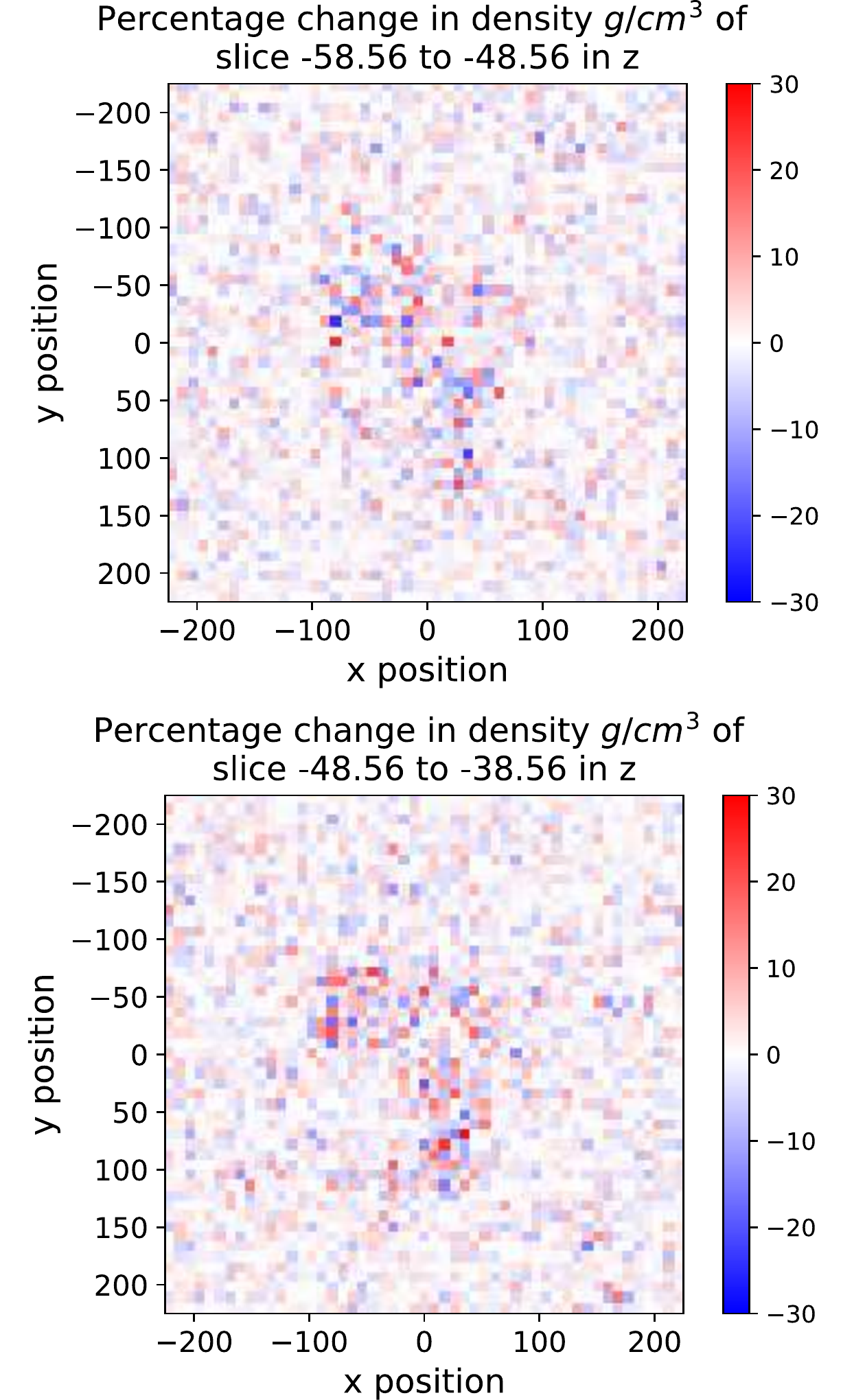}
\caption{The change in density of amorphous zircon after the 5\textsuperscript{th} collison cascade. As shown by the scale, the more red a region, the larger the increase in percentage change in density and the bluer a region the larger the decrease in percentage change in density.}
\label{fig:amorphousdensityafteroverlap4}
\end{figure}

We now look at the density variations in the amorphous structure due to collision cascades and their overlap. We investigate this effect by splitting the cell up into boxes of length 10 \si{\angstrom} and calculating the density of each box before and after each collision cascade. Examination of different slices along the z-direction of the simulation cells, as shown in Figures \ref{fig:amorphousdensityafteroverlap1} and \ref{fig:amorphousdensityafteroverlap4}, shows that radiation damage results in increased density inhomogeneities of up to 30\% of the initial regions density. The regions of density inhomogeneity grow in size with the cascade overlap as seen in Fig. \ref{fig:amorphousdensityafteroverlap4}.

Local density inhomogeneities seen as the presence of regions of increased and reduced densities (relative to the undamaged structure) have important implications for the storage of nuclear waste. Regions with decreased density serve as fast diffusion pathways in the waste form. At large radiation doses, these regions will  connect into percolating clusters where enhanced diffusion is observed on the macroscopic scale \cite{geisler_2003,Trachenko_2004per}.

We note that density inhomogeneities at the nanoscale have been reported before from small-angle x-ray scattering experiments on natural, radiation-damaged zircon \cite{rios_susana_salje_2004}, as well as from previous MD simulations of close overlaps in crystalline zircon \cite{Trachenko_2004per}. However, the effect is much more pronounced in amorphous zircon, as expected due to the increased diffusion rates of radiation damaged materials compared to their undamaged phases \cite{weber_ewing_2002}.

\begin{table}
\centering
\begin{tabular}{ccc}

Change in enthalpy & Crystal (\si{eV}) & Amorphous (\si{eV})\\
\hline

Cascade 1& \si{2822 \pm 27 } & \si{1795 \pm 25} \\
\hline
Cascade 2 & \si{2276 \pm 43 } & \si{1417 \pm 49} \\
\hline
Cascade 3 & \si{2364 \pm 37} & \si{718 \pm 53} \\
\hline
Cascade 4 & \si{1998 \pm 47} & \si{1141 \pm 50} \\
\hline
Cascade 5 & \si{1733 \pm 62} & \si{824 \pm 52} \\
\hline
Cascade 6 & \si{1608 \pm 68} &  \si{1284 \pm 52}    \\
\end{tabular}
\caption{Change in Enthalpy between each collision cascade in amorphous and crystalline zircon}
\label{table:enthalpytest}

\end{table}

We now address the energetic effects of collision cascades. We calculate the enthalpy, a macroscopically measurable quantity that can be directly compared with experiment. The enthalpy of formation of the amorphous phase compared to the crystal phase of zircon is calculated from our simulations to be 109 \si{kJ.mole^{-1}}. This can be compared with experimentally measured values of metamict zircon, natural samples of zircon with partial radiation damage over hundreds of millions of years, where the enthalpy is 59 \si{kJ.mole^{-1}}\cite{Ellsworth1994}. This is a similar order of magnitude, with potential differences attributable to the variations in structure between a fully amorphized sample in our simulations and a partially irradiated sample which has also undergone relaxation, as well as the effects of the specific interatomic potentials used.

More specifically, we can quantify the change in enthalpy between each collision cascade in both the crystal and amorphous systems, as shown in TABLE \ref{table:enthalpytest}.
Interestingly, although we give the impacting atom 70 keV of kinetic energy, there is only a total change of 2-3 keV in the enthalpy of the crystalline system. An even smaller change in the enthalpy of the amorphous system is observed, 0.8-1.7 keV per cascade. The difference in the enthalpy changes between the two systems can be explained by the difference in the initial entropy of the crystalline system, which is in the global minimum of free energy, compared to the amorphous system. In crystalline zircon radiation damage is producing some coordination changes, but most of the entropy change is attributed to amorphizing the connectivity of the crystalline structural units, rather than breaking them. In the amorphous structure the majority of the change in structural entropy comes from the breaking of the Si-O, and to a lesser extent the Zr-O, structural units, as the system is already fully amorphized.

\begin{figure}
\centering
\includegraphics[width = 8cm]{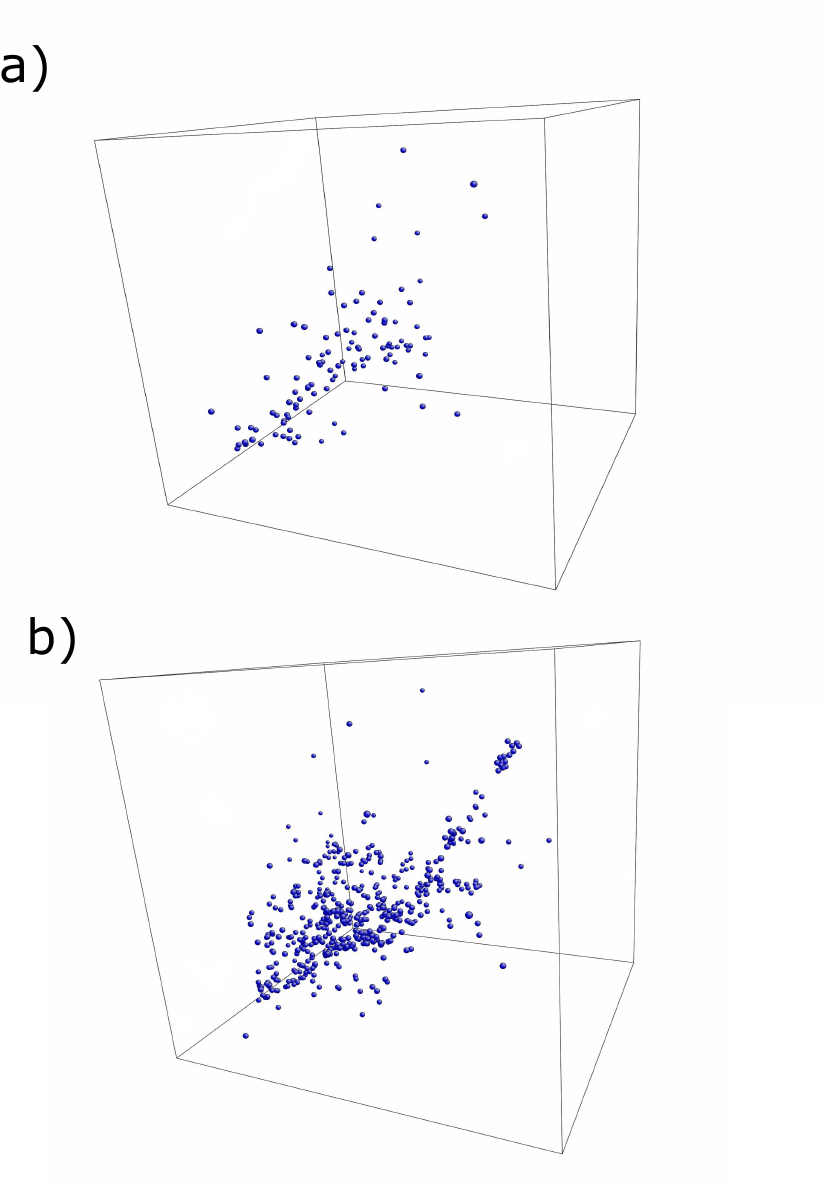}
\caption{Positions of all newly created 3 coordinated Si-O atoms inside the simulation cell for a) crystalline and b) amorphous zircon after 6 overlapping collision cascades.}
\label{fig:noframenew3coordSi}
\end{figure}

The effect of the difference between breaking structural units and amorphizing the structure is shown in Fig. \ref{fig:noframenew3coordSi}. We visualize the position of all the newly created 3-fold coordinated Si atoms as a result of 6 overlapping cascades in the amorphous and crystalline zircon structure. In the crystal, 160 new 3-coordinated Si atoms are created and are predominantly concentrated in the region along the initial trajectory of the U recoil nucleus. In the amorphous structure, 640 3-coordinated Si atoms are produced, suggesting that the lack of crystal lattice inhibits the rate of recovery of damaged Si-O tetrahedra. This shows that in the crystalline case the cascades mainly contribute to amorphizing the structure, while producing some breaking of structural units, but in the amorphous structure cascades instead lead to an increase in the breaking of network structural units. Even in the amorphous case the density of 3-coordinated Si is not high, demonstrating the difficulty in creating a fully saturated damaged region in the material.

\begin{figure}
\centering
\includegraphics[width = 8cm]{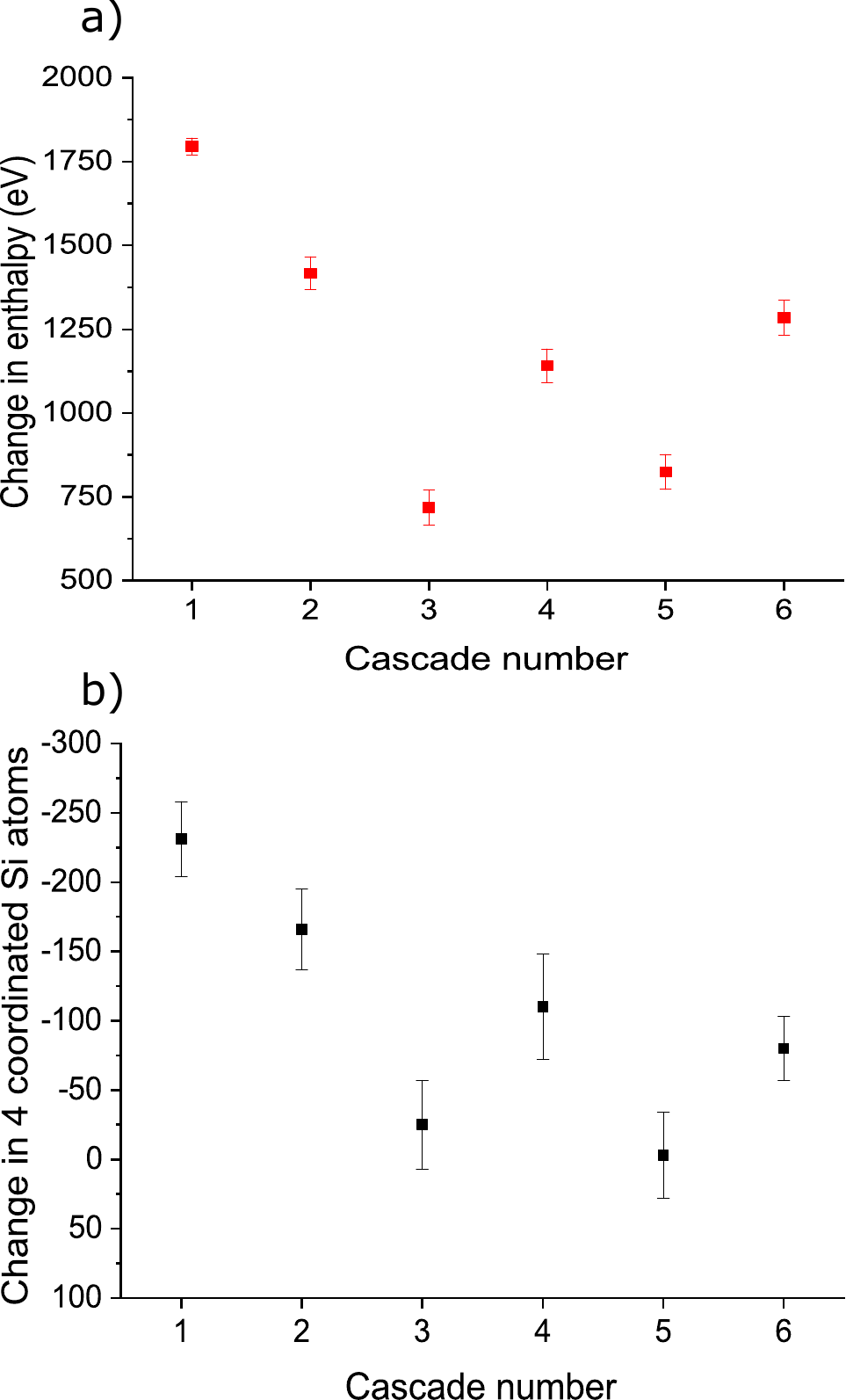}
\caption{Change in both a) the number of 4 coordinated Si-O and b) the system enthalpy between each collision cascade in amorphous zircon.}
\label{fig:dif4Si-O}
\end{figure}

Interestingly, we observe that both the changes in the number of 4-coordinated Si-O and the enthalpy of the amorphous structure (see Fig. \ref{fig:dif4Si-O}) are highly correlated between subsequent cascades, adding further evidence that the enthalpy change in amorphous zircon is linked to the breaking of Si-O bonds.  Further, both enthalpy and Si-O coordination are measurable using experimental techniques which would allow direct verification of this prediction. This implies that the macroscopic change in enthalpy could be directly linked to an atomic level change in the zircon structure.

An important conclusion from our analysis is that the local structures in the amorphous system substantially {\it evolve} with each collision cascade but these changes cannot be observed in the pair distribution functions of the material where minimal to no variation is observed. This result implies that amorphous structures can significantly evolve at the local structural level. Apart from the insight into the nature of the amorphous state itself \cite{zallen,wrightobservations,wright2013}, this has important consequences for practical applications, such as nuclear waste storage due to density inhomogeneities. 

We have not reached a point in our simulations where the amorphous structure saturates or converges to an amorphous structure which no longer changes in response to radiation damage, measured using local coordination numbers. This is due to limitations related to computational costs. However, one expects that such a convergence eventually takes places due to the need to fulfill chemical and connectivity constraints at the local level \cite{phillips}. The nature of this converged amorphous state will be of substantial interest in future experimental and modelling effort, in view of continuing changes of coordination and density at the nanoscale.

\section{Summary}

In summary, we have proposed a new method to detect and quantify damage in amorphous structures. Combining this method and large-scale molecular dynamics simulations in zircon, we have arrived at several insights regarding the response of crystalline and amorphous structures to radiation damage. Whereas an average, macroscopic property of the material, such as the pair distribution function, does not detect the differences between the damaged configurations in the amorphous structures, our new analysis method based on local atomic coordination finds that the amorphous structure substantially {\it evolves} as a result of radiation damage. Our analysis also shows a notable correlation between the local structural changes and changes in enthalpy. Importantly for the long-term immobilization of nuclear waste, we find significant local density inhomogeneities in the amorphous structure due to radiation damage. Although we do not reach the point of convergence where the changes of the amorphous structure saturate, we have proposed that our results open up the way to study this process and imply that ascertaining the nature of this new converged amorphous state will be of substantial interest in future experimental and modelling work.

\section{Acknowledgements}

AD, OD and KT were supported by the UK Engineering and Physical Sciences Research Council (EPSRC) grant EP/R004870/1.

Via our membership of the UK's HEC Materials Chemistry Consortium, which is funded by EPSRC (EP/L000202, EP/R029431), this work used the ARCHER UK National Supercomputing Service (http://www.archer.ac.uk) and the UK Materials and Molecular Modelling Hub for computational resources, MMM Hub, which is partially funded by EPSRC (EP/P020194).

Part of this work made use of computational support by CoSeC, the Computational Science Centre for Research Communities, through CCP5: The Computer Simulation of Condensed Phases, EPSRC grant no EP/M022617/1 through AME and ITT.

\bibliographystyle{apsrev4-2}
\bibliography{Zirconref}

\end{document}